\documentclass[preprint,aps,superscriptaddress]{revtex4}
\usepackage[normalem]{ulem}
\usepackage{graphicx}
\usepackage{subfigure}
\usepackage{epsfig}
\usepackage{xcolor}
\usepackage{dcolumn}
\usepackage{bm}
\usepackage{ulem}
\usepackage{color}
\usepackage{amsmath}
\usepackage{amsfonts}
\usepackage{amssymb}
\usepackage{epstopdf}

\begin{document}

%\linenumbers

\title{Raman signatures of inversion symmetry breaking and structural phase transition  in type-II Weyl semimetal MoTe$_2$}

\author{Kenan Zhang}
\affiliation{State Key Laboratory of Low Dimensional Quantum Physics and Department of Physics,Tsinghua University, Beijing 100084, China}

\author{Changhua Bao}
\affiliation{State Key Laboratory of Low Dimensional Quantum Physics and Department of Physics,Tsinghua University, Beijing 100084, China}

\author{Qiangqiang Gu}
\affiliation{International Center for Quantum Materials, School of Physics, Peking University, Beijing 100871, China}

\author{Xiao Ren}
\affiliation{International Center for Quantum Materials, School of Physics, Peking University, Beijing 100871, China}

\author{Haoxiong Zhang}
\affiliation{State Key Laboratory of Low Dimensional Quantum Physics and Department of Physics,Tsinghua University, Beijing 100084, China}

\author{Ke Deng}
\affiliation{State Key Laboratory of Low Dimensional Quantum Physics and Department of Physics,Tsinghua University, Beijing 100084, China}

\author{Yang Wu}
\altaffiliation{Correspondence should be sent to wuyangthu@mail.tsinghua.edu.cn and syzhou@mail.tsinghua.edu.cn}
\affiliation{State Key Laboratory of Low Dimensional Quantum Physics and Department of Physics,Tsinghua University, Beijing 100084, China}
\affiliation{Tsinghua-Foxconn Nanotechnology Research Center, Tsinghua University, Beijing 100084, China}

\author{Yuan Li}
\affiliation{International Center for Quantum Materials, School of Physics, Peking University, Beijing 100871, China}
\affiliation{Collaborative Innovation Center of Quantum Matter, Beijing, China}

\author{Ji Feng}
\affiliation{International Center for Quantum Materials, School of Physics, Peking University, Beijing 100871, China}
\affiliation{Collaborative Innovation Center of Quantum Matter, Beijing, China}

\author{Shuyun Zhou}
\altaffiliation{Correspondence should be sent to wuyangthu@mail.tsinghua.edu.cn and syzhou@mail.tsinghua.edu.cn}
\affiliation{State Key Laboratory of Low Dimensional Quantum Physics and Department of Physics,Tsinghua University, Beijing 100084, China}
\affiliation{Collaborative Innovation Center of Quantum Matter, Beijing, China}

\begin{abstract}
{\bf Transition metal dichalcogenide MoTe$_2$ is an important candidate for realizing the newly predicted type-II Weyl fermions, for which the breaking of the inversion symmetry is a prerequisite. Here we present direct spectroscopic evidence for the inversion symmetry breaking in the low temperature phase of MoTe$_2$ by systematic Raman experiments and first principles calculations. We identify five lattice vibrational modes which are Raman active only in the low temperature noncentrosymmetric structure. A hysteresis is also observed in the peak intensity of inversion symmetry activated Raman modes, confirming a temperature induced structural phase transition with a concomitant change in the inversion symmetry. Our results provide definitive evidence for the low temperature noncentrosymmetric \emph{T}$_d$ phase from vibrational spectroscopy, and suggest MoTe$_2$ as an ideal candidate for investigating the temperature induced topological phase transition.}
\end{abstract}

\maketitle

Layered transition metal dichalcogenides (TMDs) have attracted extensive research interests due to their intriguing physical properties for both fundamental research and potential applications in electronics, optoelectronics, spintronics and valleytronics \cite{StranoRev, ZhangHRev}. So far most of the research has been focused on semiconducting TMDs with hexagonal or trigonal (2H or 1T) structures which show strong quantum confinement effects in atomically thin films. In recent years, TMDs with monoclinic 1\emph{T}$^\prime$ and orthorhombic \emph{T}$_d$ phase have been proposed to be important host materials for realizing novel topological quantum phenomena, e.g. quantum spin Hall effect \cite{FuLSci2014, zheng2016quantum} and Weyl fermions \cite{Soluyanov2015Type}. Weyl fermions were originally introduced in high energy physics by Hermann Weyl \cite{Weyl1929}, and their condensed matter physics counterparts have not been realized until recently in Weyl semimetals in the TaAs family \cite{Xu2015TOPOLOGICAL, Lv2015Observation, Lv2015Experimental}. Weyl fermions can be realized by breaking either the time reversal symmetry or inversion symmetry of a three dimensional Dirac fermion such that a pair of degenerate Dirac points separate into two bulk Weyl points with opposite chiralities, which are connected by topological Fermi arcs when projected onto the surface.  Recently, it has been predicted that a new type of Weyl fermions can be realized in TMDs. Different from type-I Weyl fermions which have point-like Fermi surface and obey Lorentz invariance, the newly predicted type-II Weyl fermions emerge at the topological protected touching points of an electron and a hole pocket with strongly tilted Weyl cones \cite{Soluyanov2015Type}.  Such type-II Weyl fermions  break Lorentz invariance and therefore do not have counterparts in high energy physics.

\begin{figure}
  \centering
  \includegraphics[width=12cm]{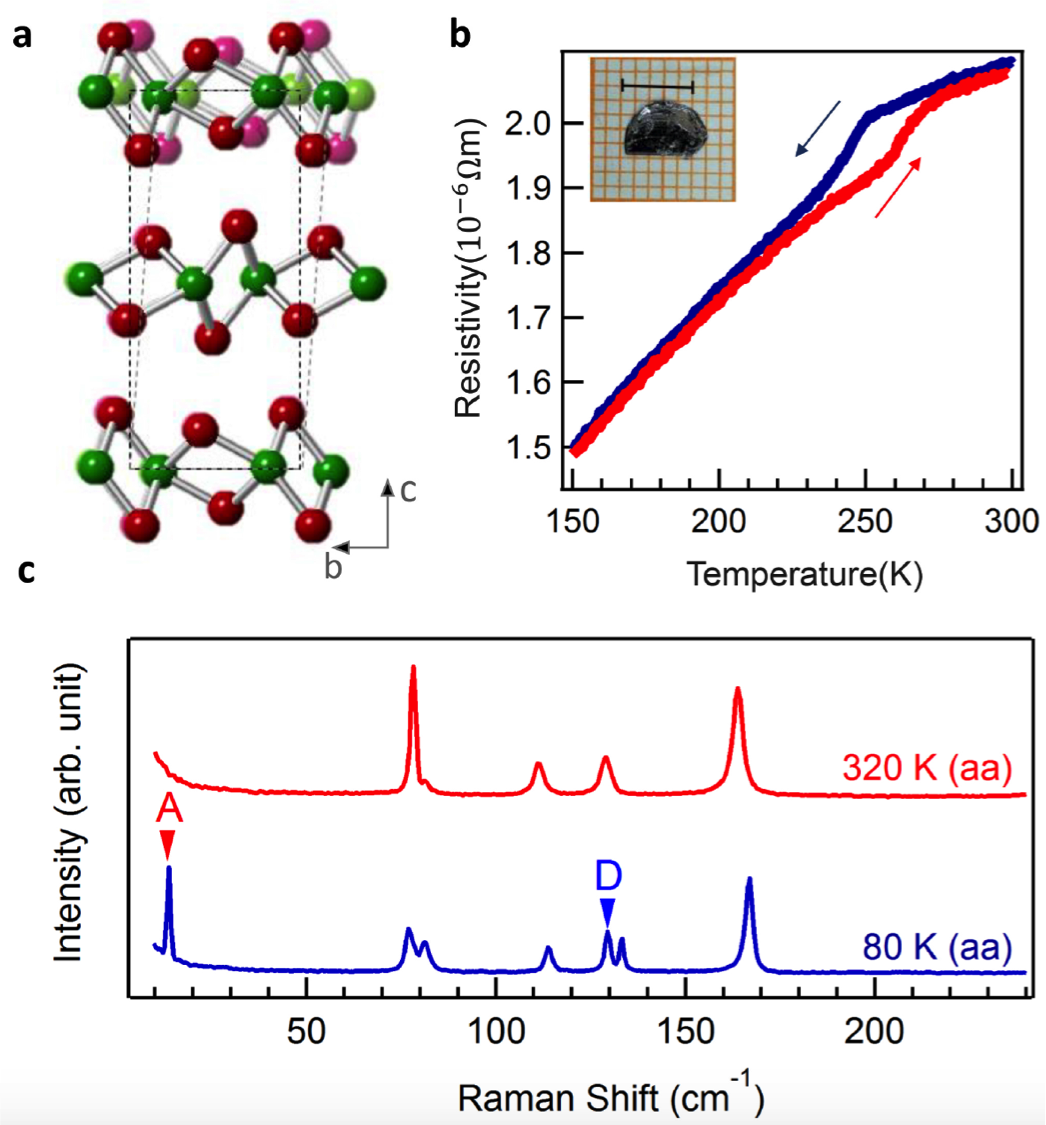}
  \caption{{\bf Temperature induced phase transition in MoTe$_2$.} ({\bf a}) Crystal structures of 1\emph{T}$^\prime$ (shadow) and \emph{T}$_d$ (solid) phases.  ({\bf b}) Resistivity measurement shows a temperature induced phase transition. The inset shows a photograph of the high quality single crystal and the scale bar is 5mm. ({\bf c}) Raman spectra at 320 K and 80 K respectively.  The letters inside the parenthesis indicate the polarization directions for incident and scattering lights.} \label{Fig1}
\end{figure}

\begin{figure}
  \centering
  \includegraphics[width=16.8cm]{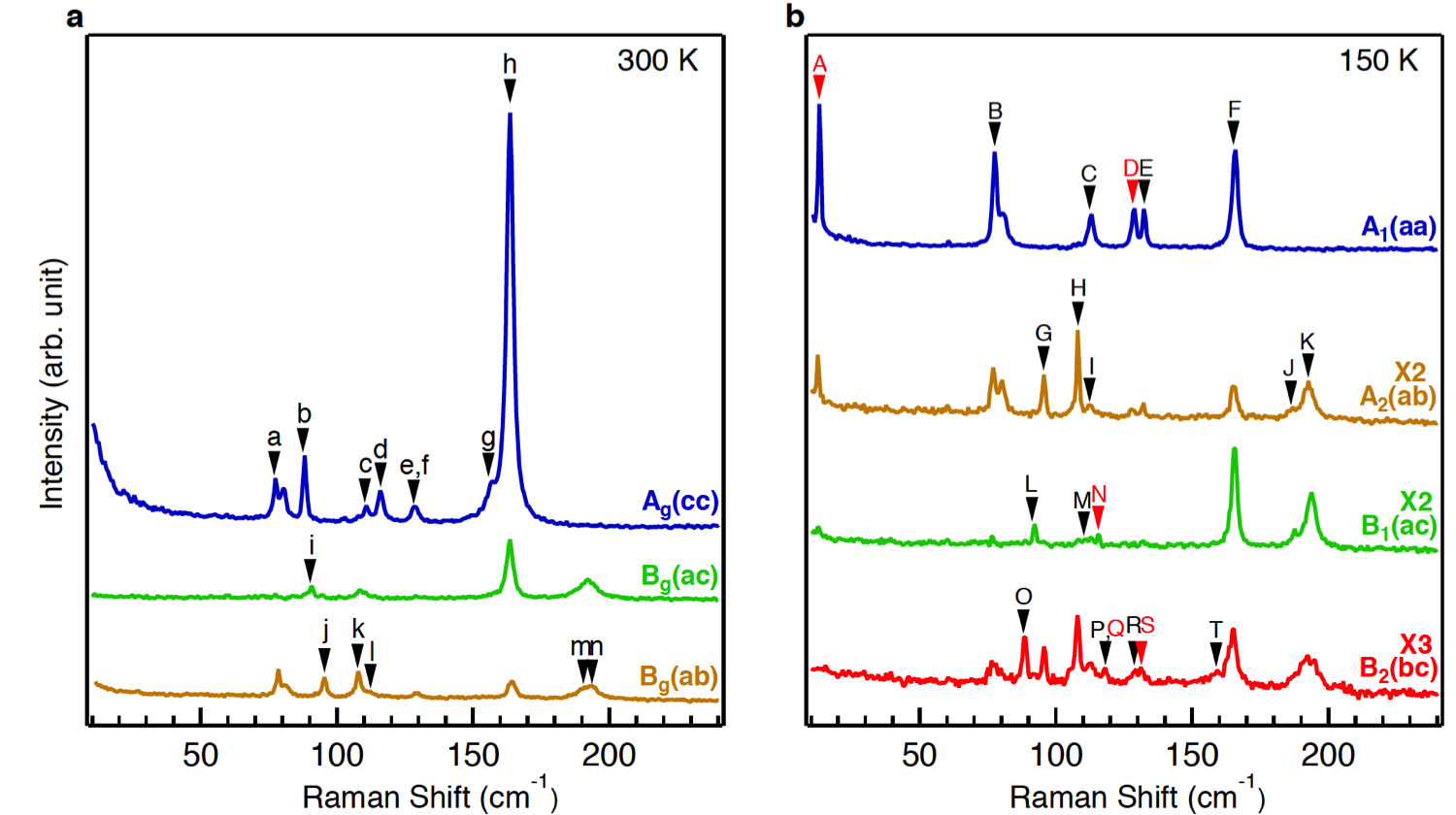}
  \caption{{\bf Polarized Raman spectra measured in the high and low temperature phases.} ({\bf a, b}) Polarized Raman spectra measured at 300 K (a) and 150 K  (b) respectively.  The identified Raman peaks are labeled by lowercase (high temperature phase) and capital (Low temperature phase) letters. The red labels A, D, N, Q and S mark the Raman modes that are directly sensitive to the inversion symmetry breaking. The small peak at 83 cm$^{-1}$ is from the instrument and not intrinsic to the sample.}\label{Fig2}
\end{figure}

 Type-II Weyl fermions have been first predicted in the orthorhombic (\emph{T}$_d$) phase of WTe$_2$ with space group \emph{Pmn}2$_1$  \cite{Soluyanov2015Type}. However, it is challenging to observe the extremely small Fermi arcs in WTe$_2$ due to the small separation of the Weyl points (0.7\% of the Brillouin zone).  Weyl fermions have also been predicted in the low temperature phase of MoTe$_2$ with much larger Fermi arcs \cite {Sun2015Prediction, Wang2015MoTe2}, and signatures of the Fermi arcs have been suggested in a combined angle-resolved photoemission spectroscopy (ARPES) and scanning tunneling spectroscopy (STS) study \cite{Deng2016} and other ARPES studies \cite{KaminskiARPES, YLChenARPES, DingARPES,ZhouARPES,BaumbergerARPES}.  The existence of Weyl fermions has been relied on the assumption that the low temperature phase of MoTe$_2$ is isostructural to the noncentrosymmetric \emph{T}$_d$ phase of WTe$_2$  \cite {Sun2015Prediction, Wang2015MoTe2}. The high temperature monoclinic 1\emph{T}$^\prime$ phase with an inclined staking angle of $\sim93.9^\circ$ has a centrosymmetric \emph{P}2$_1$/\emph{m} space group.  Although a temperature induced structural transition with a change in the stacking angle from $\sim93.9^\circ$ to 90$^\circ$ has been reported both crystallographyically   \cite{Clarke,Dawson1987} and computationally \cite{Chang2015Arc}, there are two possible space groups can be assigned to the low temperature orthorhombic phase - noncentrosymmetric \emph{Pmn}2$_1$ and centrosymmetric Pnmm  \cite{Clarke}. Previous X-ray diffraction study was limited to resolve the subtle differences between these two space groups to provide conclusive evidence on the inversion symmetry \cite{Clarke}. Recent ARPES studies have detected Fermi arcs at low temperature \emph{T}$_d$ phase, but the evolution of band dispersions from high to low temperature has not been reported. Meanwhile the absence of Fermi arcs at high temperature 1\emph{T}$^\prime$  phase are difficult to be observed because of the reduction of Resolution. So ARPES is not suitable for the research of symmetry breaking during phase transition in MoTe$_2$. Since the noncentrosymmetry is a prerequisite for realizing Weyl fermions for non-magnetic materials, it is critical to reveal the inversion symmetry breaking from spectroscopic measurements which are directly sensitive to the crystal symmetry.

 {\bf Results}

{\bf Polarized Raman spectra.} In this paper, we provide direct experimental evidence for the inversion symmetry breaking in the low temperature phase of MoTe$_2$ and study its evolution across the temperature induced structural phase transition using Raman vibrational spectroscopy.  Our Raman measurements reveal the emergence of five Raman and IR active modes in the low temperature phase, and they are in good agreement with first principles calculations and symmetry analysis of the \emph{T}$_d$ phase. These peaks are however absent in the high temperature centrosymmetric 1\emph{T}$^\prime$ phase, suggesting that they are Raman signatures for the breaking of the inversion symmetry. A clear hysteresis is observed in the peak intensity of two A$_1$ modes - the shear mode at $\approx$ 13 cm$^{-1}$ and the out-of-plane vibration mode at $\approx$ 130 cm$^{-1}$. Our results provide clear evidence for the lack of inversion symmetry in the low temperature \emph{T}$_d$ phase from a lattice dynamics point of view, and indicate that MoTe$_2$ can be a good candidate for studying the temperature induced topological phase transition.

\begin{table*}
  \centering
  \caption{Comparison of the calculated and experimental Raman modes in the 1\emph{T}$^\prime$ phase and \emph{T}$_d$ phase, in units of cm$^{-1}$.}\label{Table1}
    \resizebox{\textwidth}{!}{
  \begin{tabular}{cccccccccccccc}
    \hline
    \hline
    \multicolumn{14}{c}{\textbf{1\emph{T}$^\prime$ Phase}}\\
    \hline
    \hline
    \multicolumn{14}{c}{\textbf{Raman Active}}\\
    \hline
 {\textbf{A$_g$}}
   & $\omega_{cal}$ & 78 & 89 & 114 & 119 & 133 & 134 & 157 & 166  & 240 & 252 & 268 & 271\\
   & $\omega_{exp}$  & 77 & 88 & 110.8 & 116 & 128 & 128 & 158 & 164 &  & & & \\
   & label & a & b & c & d & e & f & g & h & &  &  & \\
    \hline
    {\textbf{B$_g$}} & $\omega_{cal}$ & 93 & 98 & 112 & 115 & 200 & 204 &  &  &  &  &  & \\
    & $\omega_{exp}$ & 90 & 94 & 107 & 111.4 & 191 & 193 &  &  &  & & & \\
    & label & i & j & k & l & m & n &  &  & &  &  & \\
    \hline
    \multicolumn{14}{c} {\textbf{IR Active}}\\
    \hline
    {\textbf{A$_u$}} & $\omega_{cal}$ & 0 & 36 & 115 & 119 & 192 & 194 &  &  &  &  &  & \\
   \\
     &label   &  &  &  & {N$^\prime$} &  &  &  & &  &  & &\\
    \hline
    {\textbf{B$_u$}} & $\omega_{cal}$ & 0 & 0 & 11 & 37 & 120 & 129 & 136 & 142 & 211 & 212 &275 & 276\\
    \\
    &label &    &  & {A$^\prime$} & & {Q$^\prime$} & {D$^\prime$} & {S$^\prime$} & &  &  & &\\
    \hline
    \hline
   \end{tabular}
   ~~~
   \begin{tabular}{cccccccccccccc}
    \hline
    \hline
    \multicolumn{14}{c}{\textbf{\emph{T}$_d$ Phase}}\\
    \hline
    \hline
    \multicolumn{14}{c}{\textbf{Raman Active}}\\
    \hline
    {\textbf{A$_2$}} & $\omega_{cal}$ & 36 & 98 & 112 & 115 & 192 & 200 &  &  &  &  &  & \\
   & $\omega_{exp}$ &  & 96 & 108 & 112.0 & 188 & 194 &  &  &  & & & \\
   & label &  & G & H & I & J & K &  &  & &  &  & \\
    \hline

        \multicolumn{13}{c}{\textbf{Raman \textnormal{\&} IR Active}}\\
    \hline
     {\textbf{A$_1$}} & $\omega_{cal}$ &0& 14 & 78 & 115 & 129 & 133 & 142 & 165 & 211 & 248 & 267 & 276 \\
    &  $\omega_{exp}$ && 13 & 77 & 112.5 & 128 & 132 &  & 165 &  &  & &  \\
    & label & & A & B & C & D & E &  & F &  & &  &   \\
    \hline
    {\textbf{B$_1$}} & $\omega_{cal}$ & 0 & 93 & 115 & 119 & 194 & 205 &  &  &  &  &  & \\
    & $\omega_{exp}$ &  & 92 & 111 & 115 &  &  &  &  &  & & & \\
    & label &  & L & M & N &  &  &  &  & &  &  & \\
    \hline
    {\textbf{B$_2$}} & $\omega_{cal}$ & 0 & 37 & 89 & 119 & 121 & 134 & 136 & 159 & 211 & 248 &270 & 277\\
    & $\omega_{exp}$ &  &  & 88 & 118 & 118 & 129 & 131 & 159 &  &  &  & \\
    & label &  &  & O & P & Q & R & S & T & &  &  & \\
    \hline
    \hline
  \end{tabular}
  }
\end{table*}

Figure \ref{Fig1}(a) shows a comparison of the low temperature (solid) and high temperature (shadow) phases with corresponding space groups of \emph{Pmn}2$_1$ and \emph{P}2$_1/$\emph{m }respectively. They share almost the same in-plane crystal structure with zigzag Mo metal chains and distorted Te octahedra.  The structural phase transition is revealed by an anomaly in the temperature-dependent resistivity \cite{Hughes}, which occurs at $\approx$ 260 K upon warming and $\approx$ 250 K upon cooling (Fig. \ref{Fig1}(b)). Figure \ref{Fig1}(c) shows the Raman spectra at 320 K and 80 K  respectively on cleaved bulk single crystals.  The polarizations for incident and scattered photons are denoted by two letters representing the crystal axes. For example, aa shows that both the incident and scattered photons are polarized along the a-axis direction. Here, we used the crystal axes of \emph{T}$_d$ phase to denote the polarization directions and all single crystal samples were oriented using Laue diffraction patterns (see Supplementary Figure 1) before performing Raman characterizations. The comparison of Raman spectra reveals two new peaks labeled by A and D only in the low temperature \emph{T}$_d$ phase, suggesting that these Raman peaks may link with the structural phase transition.

To understand the Raman modes, we first perform group theory analysis. Both the 1\emph{T}$^\prime$ phase and \emph{T}$_d$ phase have 12 atoms in one unit cell and correspondingly, there are a total of 36 phonon modes.  The vibrational modes in the 1\emph{T}$^\prime$ phase decompose into 36 irreducible representations: [12A$_g$+6B$_g$]+[5A$_u$+10B$_u$]+[A$_u$+2B$_u$], where the first, second, and third groups of irreducible representations correspond to the Raman active, infrared (IR) active, and the acoustic modes, respectively. Since the IR active and Raman active modes are exclusive of each other in centrosymmetric structures, the IR active modes A$_u$ and B$_u$ cannot be observed in Raman measurements. In the T$_d$ phase, the vibration modes decompose into 36 irreducible representations: [11A$_1$+6A$_2$+5B$_1$+11B$_2$]+[11A$_1$+5B$_1$+11B$_2$]+[A$_1$+B$_1$+B$_2$], where A$_1$, B$_1$, and B$_2$ modes are both IR and Raman active while A$_2$ modes are only Raman active.

\begin{figure*}
  \centering
  \includegraphics[width=16cm]{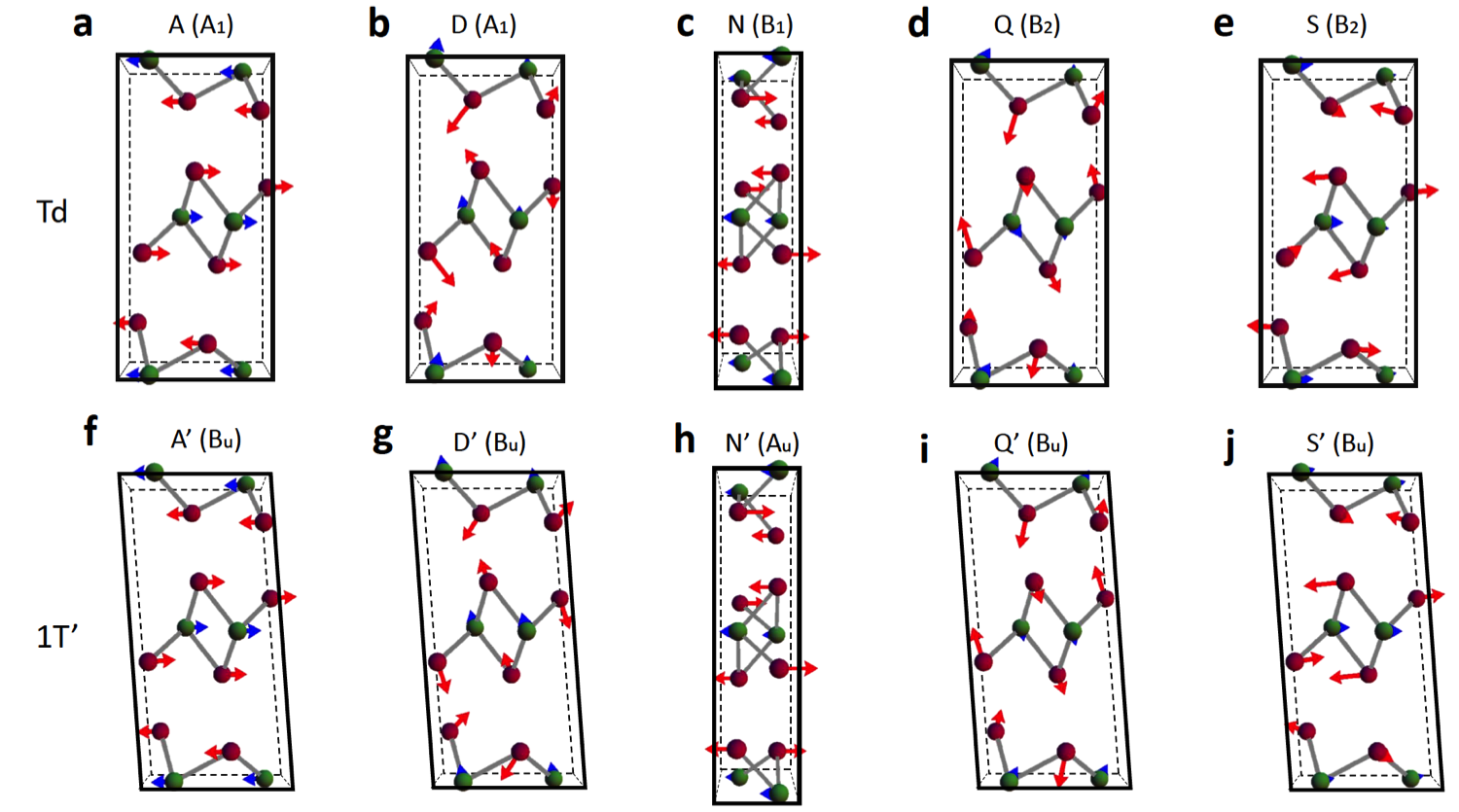}
  \caption{{\bf Calculated vibrational patterns for Raman modes that are directly sensitive to the inversion symmetry breaking.} Calculated vibrational patterns in the \emph{T}$_d$ phase ({\bf a-e}) and corresponding B$_u$ and A$_u$ modes in the 1\emph{T}$^\prime$ phases ({\bf f-j}). A, D, Q, S modes vibrate in the bc plane and N mode in the ac plane.}\label{Fig3}
\end{figure*}

Figure \ref{Fig2} shows an overview of the polarized Raman spectra measured at 300 K and 150 K.  Raman selection rules (see Supplementary Table 1, and Table 2) indicate that the A$_g$ modes in 1\emph{T}$^\prime$ phase can be observed in the aa, bb, cc, and bc configurations, whereas the B$_g$ modes can be observed in the ac and ab configurations. To obtain all possible phonon modes at low wave number in the 1\emph{T}$^\prime$ phase, we performed Raman measurements in the cc, ac, and ab configurations at 300 K. The azimuthal dependence of the Raman peak intensities for A$_g$ and B$_g$ modes (see Supplementary Figure 2) further confirm the good alignment \cite{Ribeiro2015Unusual,He2016Coupling}. Eight sharp peaks of pure A$_g$ modes are detected in the cc configuration and all the six B$_g$ modes are found in the ac and ab configurations. The sharp peaks observed are due to improved sample quality and more peaks can be resolved clearly. In the \emph{T}$_d$ phase, the A$_1$ modes can be observed in the aa, bb, and cc configurations, whereas the A$_2$, B$_1$, B$_2$ modes can only be observed in the ab, ac, and bc configurations, respectively.  The signal leakage of A$_1$ in other polarization configurations is likely due to the imperfect cleavage of ac and bc surfaces from plate-like samples, but this does not intervene the conclusion. In the low temperature phase, we observe six pure A$_1$ modes in the aa configuration, five A$_2$ modes in the ab configuration, three B$_1$ modes in the ac configuration, and six B$_2$ modes in the bc configuration.

{\bf signature of phase transition and symmetry breaking.} The comparison of Raman modes between experimental results and theoretical calculations in Table \ref{Table1} shows a good agreement.  Here we focus on Raman active modes that are sensitive to the breaking of the inversion symmetry across the phase transition. Owing to the crystal structure changes only slightly across the phase transition, we can track each phonon mode by comparing their vibrational pattern in these two phases. Since the breaking of inversion center, some Raman in-active modes that belong to A$_u$ or B$_u$ representations in the 1\emph{T}$^\prime$ phase evolve to A$_1$, B$_1$, or B$_2$ representations that are both IR and Raman active in the \emph{T}$_d$ phase. Thus, the presence of these Raman modes reflects the transition into the noncentrosymmetric phase. Similarly, nonlinear optical method was employed to reveal the lack of inversion symmetry in few-layer MoS$_2$ and h-BN \cite{{Kumar2013},{Li2013}}.

\begin{figure}
	\centering
	\includegraphics[width=13.5cm]{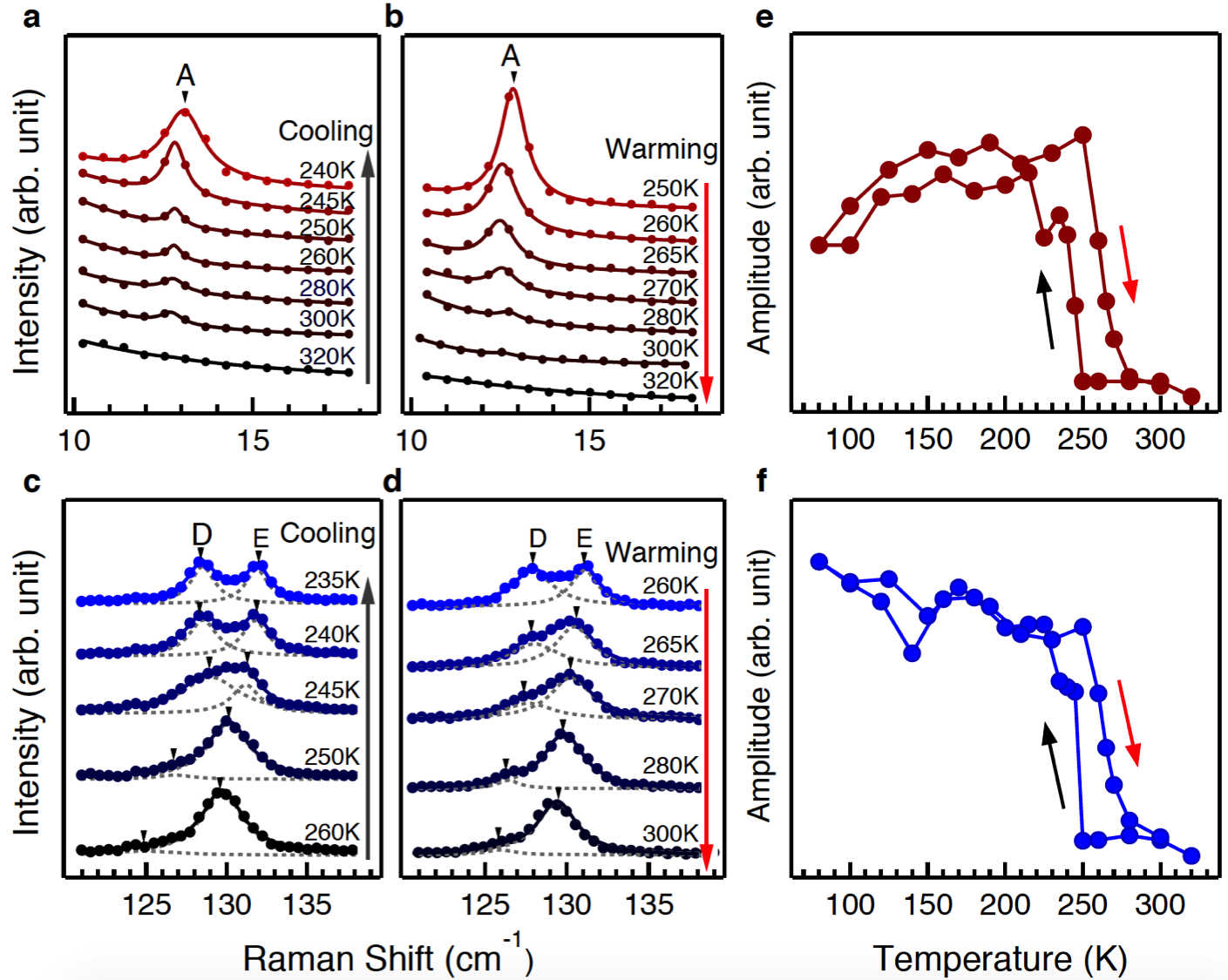}
	\caption{{\bf Temperature dependence of the Raman spectra for peaks A and D in the aa configuration.} ({\bf a-d}) Selected Raman spectra across the phase transition for peaks A (a, b) and D (c, d) upon cooling (a, c) and warming (b, d). ({\bf e, f}) Temperature dependence of the Raman intensity for peaks A and D. }
	\label{Fig4}
\end{figure}

Figure \ref{Fig3} compiles the calculated vibrational patterns for such phonons that are directly sensitive to the inversion symmetry breaking, where the arrows scale the atomic displacements. The upper panels show the vibration modes labeled by A, D, N, Q, S, which indicated by red arrows in Fig. \ref{Fig2}(b), where the irreducible representations in the \emph{T}$_d$ phase are given in parenthesis. The lower panels show the corresponding vibration modes labeled by A$^\prime$, D$^\prime$, N$^\prime$, Q$^\prime$, S$^\prime$ that belong to the A$_u$ and B$_u$ irreducible representations in 1T$^\prime$ phase. These phonons have almost identical vibrational patterns as A, D, N, Q, S respectively but no Raman activity due to the centrosymmetry. The A and A$^\prime$ denote interlayer shear modes along the b axis and the A peak is also observed in a previous report \cite{Chen2016}.  The absence of the vibrational mode along the a axis (labeled as S$_2$ in Ref.~\onlinecite{Chen2016}) in our measurements is likely due to Raman scattering matrix element effects or small Raman cross-section. Compared to previous work, here we present a systematic Raman characterization of the low temperature phase by distinguishing all modes that reflect the breaking of the centrosymmetry. The strongest Raman signals that distinguished these two phases are the interlayer shear mode A at $\approx$ 13 cm$^{-1}$ and another out-of-plane vibration mode D at $\approx$ 130 cm$^{-1}$. Similar vibrational modes have been reported in many 2D materials, such as multilayer graphene \cite{TanNatureMater,Lui2012Observation,Lui2013Measurement} and TMDs e.g. MoS$_2$, WSe$_2$ \cite{TanChemSocRev,Zhao2013,Zhao2013Lattice,Zhang2013Raman}. The low frequancy interlayer shear modes are sensitive to the stacking sequence, layer number, and symmetry, and can be used as a measure of interlayer coupling. In in-plane shear modes all atoms in the same layer all vibrates along the same direction while atoms in two adjacent layers vibrates along opposite directions. If there is an inversion center that lies in the layer, such shear modes have odd parity with respect to the inversion symmetry and therefore are Raman inactive. This is the reason why A mode is invisible in 1\emph{T}$^\prime$ MoTe$_2$, opposite to high symmetric 2H-MoS$_2$, MoSe$_2$, WSe$_2$, and 2H-MoTe$_2$. However, when the crystal structure does not hold inversion symmetry, these modes are both Raman and IR active and visible in Raman spectroscopy, giving direct evidence on the breaking of centrosymmetry in the orthorhombic \emph{T}$_d$ structure.

We further track the evolution of peaks A and D that link with the inversion symmetry breaking across the phase transition. The evolution of the A peak at 12.5 cm$^{-1}$,  and D peak at 128.3 cm$^{-1}$ are displayed in Fig.~\ref{Fig4}(a-d). Upon warming, the intensity of the A and D peaks decreases with the sharpest decrease at $\approx$ 260 K  and eventually disappears above 300 K. Upon cooling, the A and D peaks appears at a lower temperature, and their intensities sharply increases at $\approx$ 250 K, and reach the maximum below 200 K. The intensity of the A and D peaks as a function of temperature is shown in Fig.~\ref{Fig4}(e, f). The temperature-dependent peak position and FWHM are shown in Supplementary Figure 3. A discontinuity in the temperature dependent peak position can be regarded another signature of structural phase transition of MoTe$_2$. The thermal hysteresis effect in the peak intensity is consistent with our transport measurement, confirming that these peaks directly indicate the structural phase transition from high temperature 1\emph{T}$^\prime$ to low temperature \emph{T}$_d$ phase.

{\bf Discussion}

To summarize, by performing a systematic Raman study using polarization selection rules combined with theoretical calculation, we reveal the Raman signatures of structural phase transition across the 1\emph{T}$^\prime$ to \emph{T}$_d$ phase transition and provided unambiguous evidence on the absence of inversion symmetry of \emph{T}$_d$ phase. Our work demonstrates that \emph{T}$_d$ phase of MoTe$_2$ is a strong candidate for both type-II Weyl semimetal and investigating the temperature induced topological phase transition.

{\bf Methods}

{\bf Sample growth and Raman measurement.} Single crystals of MoTe$_2$  were grown by chemical vapor transport method as reported previously \cite{Deng2016}. Raman scattering experiments were performed in a confocal back-scattering geometry on freshly cleaved single crystal surfaces along the ab, ac, and bc planes.  Parallel and cross polarizations between the incident and scattered lights were used.  Raman spectra were measured using a Horiba Jobin Yvon LabRAM HR Evolution spectrometer with the $\lambda$ = 514 nm excitation source from an Ar laser and a 1800 gr/mm grating.  A liquid nitrogen-cooled CCD detector and BragGrate notch filters allow for measurements at low wave numbers. The temperature of the sample was controlled by a liquid-nitrogen flow cryostat and a heater in a  chamber with a vacuum better than $5\times10^{-7}$ Torr.

{\bf First-principles calculations.} To determine the phonon frequencies, we performed first-principles calculations of the phonon modes at the zone center using the Vienna abinitio simulation package (VASP) \cite{Kresse1996} with the local density approximation (LDA) \cite{Perdew1981} and the projector augmented wave potentials (PAW) \cite{Kresse1999}. We set a $4\times8\times2$ Monkhorst-Pack k-point mesh and 400 eV cutoff for plane waves. The coordinates and the cell shape in Ref.\onlinecite{Dawson1987} have been fully relaxed until the forces acting on the atoms are all smaller than $10^{-4}$ eV/\AA. We use the phonopy package \cite{Togo2015}, which implements the small displacement method to obtain the phonon frequencies and vibration modes at the $\Gamma$ point.

{\bf Data availability.} The data that support the plots within this paper and other findings of this study are available from the corresponding author upon reasonable request.

\bibliographystyle{naturemag}
\bibliography{reference}

{\bf Acknowledgements}

This work is supported by the National Natural Science Foundation of China (grant No.~11274191, 11334006), Ministry of Science and Technology of China (No.~2015CB921001 and 2012CB932301).

{\bf Author contributions}

S.Z. and Y.W. conceived the research project. K.Z., H.Z. and K.D. grew and characterized the samples under supervision of Y.W.. K.Z., C.B., X.R. and Y.L. performed the Raman measurements and analyzed the data. Q.G. and J.F. performed the first principle calculations. K.Z., C.B., Q.G. Y.W. and S.Z. wrote the manuscript, and all authors commented on the manuscript.

{\bf Competing financial interests}
 The authors declare no competing financial interests.

\end{document}